\documentclass{aamas2013}
\usepackage{times}
\usepackage{latexsym}
\usepackage{helvet}
\usepackage{courier}
\usepackage{longtable,epsfig,amssymb,amsmath,indentfirst,lscape}
\usepackage{algorithmic,algorithm}
\floatname{algorithm}{Procedure}

\usepackage{etoolbox}
\makeatletter \patchcmd{\maketitle}{\@copyrightspace}{}{}{}
\makeatother

\begin{document}

\title{Strategy-Proof Prediction Markets}

%\author{{\bf Paper: 156}}

%\numberofauthors{2}

\author{ \alignauthor Ayman Ghoneim\titlenote{Corresponding author. Email: ayman.ghoneim@anu.edu.au; ayman.a.sabry@gmail.com.} and Robert
C. Williamson \\
\affaddr{The Australian National University and NICTA\\
Canberra ACT, Australia.\\}}

%\email{ayman.ghoneim@anu.edu.au,
%bob.williamson@anu.edu.au} }

%\date{April 22, 2009}

\maketitle

\begin{abstract}

Prediction markets aggregate agents' beliefs regarding a future
event, where each agent is paid based on the accuracy of its
reported belief when compared to the realized outcome. Agents may
strategically manipulate the market (e.g., delay reporting, make
false reports) aiming for higher expected payments, and hence the
accuracy of the market's aggregated information will be in question.
In this study, we present a general belief model that captures how
agents influence each other beliefs, and show that there are three
necessary and sufficient conditions for agents to behave truthfully
in scoring rule based markets (SRMs). Given that these conditions
are restrictive and difficult to satisfy in real-life, we present
novel strategy-proof SRMs where agents are truthful while dismissing
all these conditions. Although achieving such a strong form of
truthfulness increases the worst-case loss in the new markets, we
show that this is the minimum loss required to dismiss these
conditions.

\end{abstract}

% fill this part

% Note that the category section should be completed after reference to the ACM Computing Classification Scheme available at
% http://www.acm.org/about/class/1998/.


%A category including the fourth, optional field follows...
%\category{D.2.8}{Software Engineering}{Metrics}[complexity measures, performance measures]

%General terms should be selected from the following 16 terms: Algorithms, Management, Measurement, Documentation, Performance, Design, Economics, Reliability, Experimentation, Security, Human Factors, Standardization, Languages, Theory, Legal Aspects, Verification.


%Keywords are your own choice of terms you would like the paper to be indexed by.


%\vspace{-7mm}

\section{Introduction}

Prediction markets have been used widely as a powerful tool to
elicit the beliefs of agents about a future event;
see~\cite{Wolfers2004,Wolfers2006,Pennock2007,Chen2010a}. In such
markets, an agent reports a probability distribution (i.e., an
estimate) over the set of mutually exclusive and exhaustive possible
outcomes of a future event. When the outcome of this future event is
realized, the agents are paid based on the accuracy of their reports
when compared to the realized outcome. It has been shown that
prediction markets produce better estimates for future events
compared to polls and expert opinions~\cite{Berg2008,Wolfers2006}.

Given that agents are paid based on their reports, an agent can
maximize its expected payoff by strategically manipulating the
market (e.g., delaying its report, and/or making false reports), and
here, the accuracy of the market's aggregated information will be in
question. This fact highlights the fundamental relation between
prediction markets and mechanism
design~\cite{Vincent2009,Chen2010a}.

In mechanism design~\cite{Colell1995}, eliciting the private
information of agents is required to determine an outcome that
reflects their conflicting interests, where an agent's private
information defines its value of each possible outcome for the
problem. However, a prediction market problem differs slightly from
a mechanism design problem in the sense that eliciting the agents'
private beliefs is the end goal and no outcome will be determined.
The prediction market problem is more close to an
\emph{interdependent valuations} mechanism design
problem~\cite{Mezzetti2004} (i.e., an agent's value of an outcome
depends on the private information of other agents in addition to
its private information) than a classical mechanism design problem
(i.e., an agent's value of an outcome depends only on its private
information), since a realistic model for the prediction market
problem -- as the one we consider here -- should assume that an
agent's belief (therefore its report and expected payoff) is
influenced by other agents. Without such influence, scoring rule
based markets~\footnote{Also known as market scoring rules (MSR). We
use the ``scoring rule based markets" terminology since it is more
expressive.} (SRMs)~\cite{Hanson2003,Hanson2007} are merely
considering the report of the last participating agent, and may not
converge to a final estimate that encapsulates the wisdom of the
crowd.

In both mechanism design problems and prediction markets,
truthfulness is achieved under a game-theoretic solution concept
(i.e., a truth-telling equilibrium) such as \emph{dominant strategy}
(i.e., \emph{strategy-proof}) which is the strongest form of
truthfulness where an agent will be truthful even if other agents
are not, or \emph{ex-post incentive compatibility} which is a weaker
form of truthfulness where an agent will be truthful if and only if
all other agents are truthful. Unlike mechanism design problems,
prediction markets normally operate at a loss that is considered the
price for aggregating the agents' beliefs. Several studies have
addressed the strategic behavior of agents in prediction markets.
They can be categorized as follows:

%\vspace{-4mm}

\begin{enumerate}

\item Adopting a game theoretic perspective that views the market
as a game and investigates its truth-telling equilibrium under
different models, such as conditionally dependent or independent
beliefs of
agents~\cite{Bluffing,Dimitrov2008,Chen2009,Ostrovsky2001}, betting
games~\cite{Niklova2007} and decision making markets where an
outcome will be determined based on the aggregated
information~\cite{Chen2011};

%\vspace{-3mm}

\item Investigating agents' strategic behavior empirically by
evaluating the effect of manipulators~\cite{Hanson2006};

%\vspace{-3mm}

\item Adopting a mechanism design framework that produces a
mechanism rather than a market where agents not only report their
beliefs but also report the reasons behind their
beliefs~\cite{Vincent2009}.

\end{enumerate}

%\vspace{-4mm}

To the best of our knowledge, there have been no attempts to design
prediction markets which maintain a stronger form of truthfulness
compared to existing markets. In this study, we define a general
belief model that captures any possible influences between the
agents either inside or outside the market. We show that there are
three sufficient and necessary conditions for traditional (i.e.,
presently known) SRMs to be truthful. These conditions are the
\emph{predefined participation order} condition (i.e., agents report
their beliefs to the market in a predefined order), the \emph{one
participation and market influence condition} (i.e., an agent can
report only once and can influence other agents' beliefs only by
this single report), and the \emph{non-negative influence condition}
(i.e., an agent assumes that being influenced by other agents'
beliefs doesn't decrease its expected value). Given that these
conditions are very restrictive and difficult to satisfy in
real-life, we present novel \emph{strategy-proof SRMs} that achieve
truthfulness while dismissing all the previous conditions. However,
there is a trade-off between achieving such a strong form of
truthfulness and the payments made to the agents, since these
strategy-proof prediction markets make additional payments compared
to traditional SRMs. We investigate dismissing each condition
separately to evaluate its contribution to the market's loss, and we
show that these losses are the minimum possible losses to relax the
previous conditions. Also, we show that our contribution can be
extended to cost function based markets (CFMs)~\cite{Chen2007}. In
the next section, we define the prediction market problem and our
belief model. In section 3, we discuss scoring rules and SRMs. In
section 4, we define solution concepts for truthfulness and the
necessary and sufficient conditions for SRMs to be truthful. In
section 5, we present the strategy-proof SRMs and extend our work to
CFMs. Section 6 concludes the study.

%% Our belief model
%To capture this assumption, we can assume that each agent receives
%two signals (i.e., estimates), a private signal and a public signal.
%The private signal reflects the agent's personal reasoning about the
%future event. While the public signal reflects what the agent
%observes about the beliefs of other agents, either by observing
%other agents' reports to the market or through interacting with
%other agents outside the market (e.g., discussions about the future
%event with family members and/or friends). After receiving both the
%private and the public signals, the agent can reach a final true
%belief about the future event using some sort of a merging
%reasoning.
%
%This influence factor adds various complexities when it comes for
%studying the strategic behaviors of participating agents, because
%the expected value of an agent not only depends on its private
%information, but as well on the private information of other agents.
%
%Interdependent valuation settings usually have very complicated
%strategic interactions between agents, and just recently, it has
%been shown in~\cite{Ghoneim2011} that designing strategy-proof
%mechanisms is possible when valuations are interdependent.

% Add more types of prediction markets

%where $\forall p \in \Delta_{N}, \sum^{N}_{i=1} p_{i} = 1$ holds.

%\vspace{-2mm}

\section{Prediction Markets}

%\vspace{-3mm}

{\bf Problem Statement.} Consider a \emph{future event} $X$ that has
a set $\Omega = \{1, \ldots, N\}$ of mutually exclusive and
exhaustive \emph{outcomes}, and a set of agents who have
\emph{privately known} beliefs regarding that future event. Each
agent's belief is a probability distribution (i.e., an estimate) $p
= [p_{1}, \ldots, p_{N}]$ over the set $\Omega$ of all possible
outcomes for the event $X$, where $p_{i}$ is the agent's probability
that the outcome $i \in \Omega$ will be realized. Let $\Delta_{N} =
\{ p \in \mathbb{R}^{N}: 0 \leq p_{i} \leq 1, \sum^{N}_{i=1} p_{i} =
1 \}$ be the probability simplex that contains all possible
probability distributions $p$. Assuming that it is required to
obtain information about the event $X$ by eliciting the agents'
beliefs regarding that event, a market maker may establish a
prediction market where agents can report their probability
estimates and get paid based on the accuracy of their reports when
compared to the realized outcome $i$ for the event $X$.

%In this study, we only consider the classical prediction market
%problem that only deals with aggregating the beliefs of agents
%regarding a future event. However, there are other several extended
%versions of the classical problem that assume, for instance, that
%agents can take actions to affect the outcome of the future
%event~\cite{Shi2009}, or that the market maker will make some
%decisions based on the aggregated information~\cite{Chen2011}.

{\bf Belief Model.} There are several ways to model agents'
reasoning in prediction markets. We propose a belief model which
focuses on how agents influence each other's beliefs. Our belief
model captures such influence by assuming that each agent receives
two signals (i.e., probability estimates). The first signal is a
\emph{private signal} $p^{prv}$, which represents the agent's own
reasoning about the future event without any external influences.
The second signal is a \emph{public signal} $p^{pub}$, which
represents the influence of other agents, either through their
reports inside the market or through announcements and discussions
outside the market. This public signal is the only way an agent is
influenced by other agents. Each agent receives a \emph{predefined}
private and public signals, and thus, the agent cannot affect their
contents. However, an agent can influence the public signals of
other agents as we will show later.

Given the agent's private and public signals, the agent needs to
form a final true belief $p$ about the future event. To do that, the
agent must decide whether to use its public signal $p^{pub}$ along
with its private signal $p^{prv}$ to produce it final belief $p$ or
to consider its private signal as its final true belief (i.e., $p =
p^{prv}$). As we will show later, this decision depends on what the
agent thinks about how its public signal $p^{pub}$ will affect its
expected payoff. Each agent has a predefined \emph{merging function}
$M: \Delta_{N} \times \Delta_{N} \rightarrow \Delta_{N}$ that the
agent will use if it decided to incorporate its public signal
$p^{pub}$ to produce its final belief $p$, i.e., the merging
function $M(p^{prv}, p^{pub}) = p$ merges the agent's private and
public signals -- in an arbitrary but predefined way -- producing
the final belief $p$. For each agent, its $p^{prv}$, $p^{pub}$, and
$M(p^{prv}, p^{pub})$ are predefined and privately known to the
agent, i.e., the agent's private type is $\langle p^{prv}, p^{pub},
M \rangle$. Each agent knows nothing about the types of other
agents, and types may differ from one agent to another. We assume
that each agent receives only one private signal and only one public
signal, and we will later discuss relaxing this assumption.

\begin{figure}[!h]
\begin{center}
 \includegraphics[width=3in,height=1.5in]{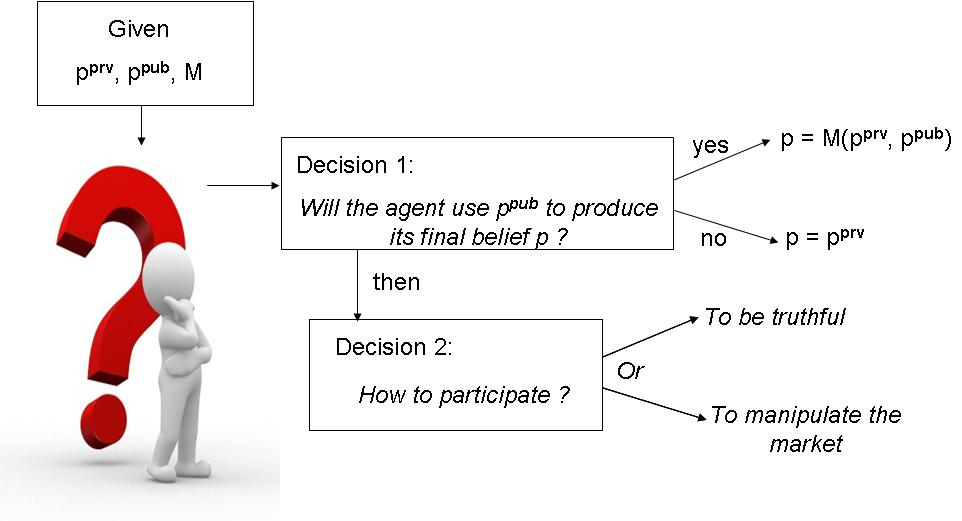}
 \caption{An agent's reasoning in the proposed belief model.} \label{fig5_1}
\end{center}
\end{figure}

Figure~\ref{fig5_1} summarizes an agent's reasoning about reaching
its final belief $p$ as we discussed earlier, and then, the agent
needs to decide whether it will participate in a truthful manner or
manipulate the market as we will discuss below.

%\vspace{-3mm}

{\bf Truthfulness and Manipulation.} Different types of prediction
markets use different protocols for how agents express their beliefs
in the market and how agents get paid. For instance, in SRMs, agents
report their beliefs directly to the market and get paid using a
particular scoring rule. While in CFMs, agents trade securities
(e.g., a ticket that pays \$1 if a particular outcome $i$ is
realized and \$0 otherwise) in the market and their beliefs are
inferred from their trading behavior. Describing a prediction market
in terms of its protocol (i.e., how it works) only provides a
partial picture of the market's dynamics, since in order to elicit
the agents' beliefs quickly and accurately other conditions (e.g.,
when and how many times agents will participate in the market) are
imposed to guarantee that agents will behave truthfully.

{\bf Definition 1.} {\it An agent \emph{behaves truthfully} if once
it receives its private signal $p^{prv}$ and public signal
$p^{pub}$, it uses its merging function $M(p^{prv}, p^{pub})$ to
produce its final belief $p$ and it reports $p$ to the market
without any delays.}

%~\footnote{Many factors can be listed under irrational behavior,
%e.g., any unintentional miscommunication that may happen when an
%agent reports its belief.}

However, an agent may manipulate the market either intentionally if
this increases its expected payoff, or due to irrational, malicious,
or any other behavior that contradicts the rationality assumption
(i.e., an agent never behaves in a way that decreases its expected
payoff). We define market manipulation as follows.

{\bf Definition 2.} {\it An agent can \emph{manipulate} the market
by: {\it 1.} delay reporting its true belief $p = M(p^{prv},
p^{pub})$; {\it 2.} reporting a false belief $p' \neq p$; and/or
{\it 3.} misleading other agents either by participating more than
once in the market and reporting false beliefs $p' \neq p$, or by
spreading false information outside the market~\footnote{We don't
detail how agents interact outside the market since how an agent is
affected by other agents is encapsulated in the public signal the
agent receives.}.}

When the market is manipulable, the accuracy of the market's
aggregated information will be in question. Similar to mechanism
design, truthfulness is achieved in prediction markets under a
game-theoretic solution concept, and we define two solution concepts
in the prediction markets context as follows.

{\bf Definition 3.} {\it A prediction market achieves truthfulness
in

\emph{Dominant Strategy (also known as strategy-proofness):} For any
rational agent, being truthful (Definition 1) always maximizes the
agent's expected payoff even if other agents are manipulating the
market (Definition 2).

\emph{Ex-Post Incentive Compatibility (also known as ex-post Nash):}
For any rational agent, being truthful (Definition 1) maximizes the
agent's expected payoff if other agents are rational and truthful
(Definition 1).}

We consider truthfulness in ex-post incentive compatibility to be an
unrealistic solution concept for real-life applications in spite of
its frequent usage in the mechanism design literature, because
assuming that an agent -- who has the ability to behave
strategically -- will assume that other agents are truthful is
unreasonable. Strategy-proofness is the strongest and most preferred
form of truthfulness, since an agent will not manipulate the market
irrespective of other agents' behavior. Moreover, achieving
strategy-proofness is motivated by the fact that it is not always
possible to assume that all involved agents are fully rational.

%Our aim in this study is to present prediction markets that operate
%while imposing less or no conditions compared to the existing
%markets. In the next section, we will demonstrate how the scoring
%rule based markets work in general, while in section 4, we will
%discuss the conditions under which the scoring rule based markets
%are not manipulable by participating agents.

%\vspace{-2mm}

\section{Scoring Rule Based Markets}

%~\cite{Hanson2003,Hanson2007}

Hanson~\cite{Hanson2003,Hanson2007} introduced SRMs as markets for
aggregating the agents' estimates, where scoring rules are used to
pay agents for their reported beliefs.

{\bf Scoring Rules.} Scoring rules have been used extensively in
aggregating and evaluating the accuracy of reported probabilistic
forecasts regarding future events~\cite{Gneiting}. A scoring rule is
a function $s: \Delta_{N} \times \Omega \rightarrow [-\infty,
\infty]$. Given a probability distribution $p \in \Delta_{N}$, a
\emph{scoring rule} $s(p) = [s_{1}(p), \ldots, s_{N}(p)]$ assigns a
score $s_{i}(p)$ that takes a value in the extended real line
$[-\infty, \infty]$ for each outcome $i \in \Omega$. The score
$s_{i}(p)$ serves as a reward (penalty) that the agent will receive
(pay) for predicting the distribution $p$, and the outcome $i$ was
realized. From now on, we will use $p$ to denote the probability
estimate that reflects the true belief of an agent, and $p'$ to
denote the probability estimate that the agent will report, which
may or may not be equivalent to $p$. Given a scoring rule $s$, an
agent's expected payoff from reporting $p'$ while having a true
belief $p$ is $\textsl{E}(s, p, p') = \sum^{N}_{i=1} p_{i}
s_{i}(p')$.

A \emph{regular} scoring rule is a scoring rule where an agent's
expected payoff $\textsl{E}(s, p, p')$ takes a value in $[-\infty,
\infty)$ for all $p, p' \in \Delta_{N}$, and the expected payoff
$\textsl{E}(s, p, p)$ for reporting the true estimate $p$ takes a
value in $(-\infty, \infty)$. This implies that $s_{i}(p)$ is finite
whenever $p_{i} > 0$. A \emph{proper} scoring rule is a scoring rule
where a risk neutral agent has no incentive to report any
distribution $p'$ other than its true belief estimate $p$, i.e.,
$\textsl{E}(s, p, p) \geq \textsl{E}(s, p, p')$, $\forall p, p' \in
\Delta_{N}$. The scoring rule is said to be \emph{strictly proper}
if the previous inequality holds with equality only when $p = p'$.
Let $b > 0$ and $a_{1}, \ldots, a_{N}$ be parameters. An example of
a regular strictly proper scoring rule is the logarithmic scoring
rule~\cite{Good}:

%\vspace{-4mm}

\begin{equation}\label{eq1}
s_{i}(p) = a_{i} + b \ln(p_{i}).
\end{equation}
%
%\noindent and the quadratic scoring rule~\cite{Brier}:
%
%\begin{equation}\label{eq2}
%s_{i}(p) = a_{i} + b (2 p_{i} - \sum^{N}_{i=1} p^{2}_{i}).
%\end{equation}

%\vspace{-1mm}

Given a proper scoring rule $s$, we can define an \emph{uncertainty
function} and a \emph{discrepancy function}~\cite{Dawid}. The
uncertainty function $S(s, x) = \sum^{N}_{i=1} x_{i} s_{i}(x)$ for
$x \in \Delta_{N}$ measures the uncertainty (i.e., lack of
precision) associated with the distribution $x$. When an agent faces
a proper scoring rule $s$, the uncertainty function $S(s, p)$ is
equal to the agent's maximum expected payoff $\textsl{E}(s, p, p)$
that results from reporting its true belief $p$. The discrepancy
function

%\vspace{-3mm}

\begin{equation}\label{eq3}
D(s, x, y) = \sum^{N}_{i=1} x_{i} s_{i}(x) - \sum^{N}_{i=1} x_{i}
s_{i}(y)
\end{equation}

%\vspace{-1mm}

\noindent for any $x, y \in \Delta_{N}$ measures the distance
between the two distributions $x$ and $y$ using the scoring rule $s$
as the metric for measurement (i.e., the distance differs
accordingly with the $s$ in use). This distance reflects the
difference between the uncertainty associated with $x$ and the
uncertainty associated with $y$ compared to $x$.

% Change the Procedures to be more conscis.
{\bf Scoring Rule Based Markets.} A scoring rule based market (SRM)
can be viewed as a sequentially shared proper scoring rule $s$ that
works as described in Procedure~\ref{Proc1}. The market always keeps
a current probability estimate $p_{c}$, which is defined by the
market maker when the market begins by an initial estimate $p_{0}$
(Step 1). Until the market closes (Steps 2-5), any agent can change
that current estimate $p_{c}$ to its reported estimate $p'$ (Step
3). The market maker saves the initial estimate and all the reported
estimates in a vector $\theta$ (Steps 1 and 4). The market's closing
(i.e. the last) current estimate represents the market's elicited
information from all the agents. Once outcome $i$ is realized, each
agent who made a report $p'$ receives $s_{i}(p')$ and pays
$s_{i}(p_{c})$, where $p_{c}$ is the current market estimate that
immediately precedes the agent's report (Steps 6-9).

%\vspace{-2mm}

\begin{algorithm}
\caption{SRM Protocol.} \label{Proc1} \algsetup{indent=2em}
\begin{algorithmic}[1]

\STATE $m \leftarrow 1$, $t \leftarrow 0$, $p_{c} \leftarrow p_{0}$
and $\theta(t) \leftarrow p_{c}$.

\REPEAT

\STATE An agent $j$ reports an estimate $p'$ and $p_{c} \leftarrow
p'$.

\STATE $t \leftarrow t + 1$ and $\theta(t) \leftarrow p'$.

\UNTIL{Market closes and outcome $i$ is realized.}

\WHILE{$m \leq t$}

\STATE $p' \leftarrow \theta(m)$, $p_{c} \leftarrow \theta(m-1)$ and
$m \leftarrow m + 1$.

\STATE The payment of agent $j$ who reported $p'$: $s_{i}(p') -
s_{i}(p_{c})$.

\ENDWHILE

\end{algorithmic}
\end{algorithm}

%\vspace{-2mm}

%We stress that in some scoring rule based markets (e.g., the
%logarithmic scoring rule based market) agents are not allowed to
%report distributions that have the probability of $0$ associated
%with any outcome $i \in \Omega$, because in this case the agent
%would have to pay (or receive) an infinite amount of money if the
%outcome with the $0$ reported probability was realized.

In a SRM that uses a proper scoring rule $s$, if an agent's true
belief estimate is $p$ and it changed the market current probability
estimate $p_{c}$ to $p'$, then its expected payoff is

%\vspace{-3mm}

\begin{equation}\label{eq4}
\textsl{E}(s, p, p', p_{c}) = \sum^{N}_{i=1} p_{i} (s_{i}(p') -
s_{i}(p_{c})).
\end{equation}

Prediction markets normally run at a loss, at least theoretically.
The market maker's loss is the price it pays to elicit the agents'
beliefs about the future event. In SRMs when outcome $i$ is
realized, each agent pays $s_{i}(p_{c})$ for the immediate previous
report $p_{c}$, which is the same amount $s_{i}(p')$ received by the
agent who reported $p' = p_{c}$. Thus, all the intermediate payments
to and from the agents offset each other, and the market maker is
left with receiving $s_{i}(p_{0})$ from the first participating
agent and paying $s_{i}(p')$ for the last participating agent who
reported $p'$. But because any outcome $i \in \Omega$ can be
realized, and the last report $p'$ can be any estimate in the
simplex $\Delta_{N}$, the market maker's \emph{worst case loss}
(WCL) is

%\vspace{-3mm}

\begin{equation}\label{eq5}
\max_{i \in \Omega} \sup_{p \in \Delta_{N}} (s_{i}(p) -
s_{i}(p_{0})),
\end{equation}

\noindent where $p_{0}$ is the initial estimate. In Eq.~\ref{eq5},
$s_{i}(p) - s_{i}(p_{0})$ reflects the highest possible price that
the market maker will pay for its additional gained knowledge
compared to its initial estimate $p_{0}$, given any possible outcome
$i$.

{\bf Example 1.} A software house wants to predict the probable
release date of a certain software, either in March (outcome A) or
in June (outcome B), and will establish a logarithmic SRM (LSRM)
that uses a logarithmic scoring rule (Eq. 1) with $a_{i} = 0,
\forall i \in \Omega$ and $b = 1$. The market starts by an initial
estimate $p_{0} = [p_{A} p_{B}] =$ [0.5 0.5] where $p_{A}$ and
$p_{B}$ are the probabilities that outcomes $A$ and $B$ will be
realized, respectively. To illustrate how this LSRM works,
Table~\ref{tab1} shows the participating agents (Column 1), and for
each agent the table shows its true belief $p$ (Column 2), its
participation order (PO) (Column 3), the market's current estimate
$p_{c}$ when it participated (Column 4), its reported belief $p'$
(Column 5), and its expected payoff (EP) (Column 6). According to
Table~\ref{tab1}, we have agent 1 and agent 2, and both are truthful
(i.e., reported $p' = p$). Agent 1 participated first, and changed
the current market estimate $p_{c} = p_{0} =$ [0.5 0.5] to $p' =$
[0.4 0.6], and its expected payoff~\footnote{The expected payoff is
approximated to four-decimals precision and is calculated using Eq.4
given $p$, $p'$ and $p_{c}$: $(0.4) [\ln(0.4) - \ln(0.5)] + (0.6)
[\ln(0.6) - \ln(0.5)] = 0.0201$.} will be $0.0201$. Then, agent 2
participated and changed the current probability $p_{c} =$ [0.4 0.6]
to $p' =$ [0.7 0.3], and its expected payoff will be $0.1838$.

%$(0.7) [ln(0.7) - ln(0.4)] + (0.3) [ln(0.3) - ln(0.6)] = 0.1838$.

Agent 2 is a programmer who can't affect the software's release
date, and we will illustrate how he came up with his final belief
$p$. Once he heard about this market, he formed his own belief
$p^{prv} = $ [0.8 0.2]. However, after a discussion with colleagues
over lunch, he became aware of some technical problems facing the
development team, and he knew that the project manager responsible
for this software was on a sick leave. Furthermore, the programmer
observed a previous report  [0.4 0.6] made to the market (i.e.,
agent 1). To capture all these external effects -- other than the
agent's initial belief $p^{prv}$, we assume that agent 2 receives a
public signal $p^{pub} =$ [0.45 0.55]. We \emph{stress} that agent 2
can't affect the content of its public signal, e.g., the agent can't
choose which colleagues to talk to or which information it receives.
However, an agent can affect the public signals of other agents,
e.g., the report of agent 1 affected the public signal of agent $2$.
We assume here that agent 2 thinks that using $p^{pub}$ to produce
its final belief will increase its expected value, and it will use a
predefined merging function to come up with its final belief $p$.
Agent 2 used a simple merging function $M'(p^{prv}, p^{pub})$ that
works for an event with two outcomes as follows: the final belief
$p$ is $p^{prv}$ after increasing the probability of the outcome
that had a higher probability in $p^{pub}$ by $0.1$, and decreasing
the the probability of the other outcome by $0.1$. Given that
outcome B (i.e., software released in June) has a higher probability
in $p^{pub}$, then $p =$ [(0.8-0.1) (0.2+0.1)] = [0.7 0.3] (i.e.,
the true belief $p$ of agent 2 in Table~\ref{tab1}). It is clear
here that the reported belief of agent 2 is the final current
estimate in the market which reflects the elicited beliefs of all
agents. If agents don't influence the beliefs of each other (e.g.,
agent 2 wasn't influenced by the report of agent 1), then the market
would only have elicited the belief of the last participating agent.

%\vspace{-3mm}

\begin{table}[!h]
\caption{Example 1}
\begin{center}
\label{tab1} \footnotesize
\begin{tabular}{|c|c|c|c|c|c|}

\hline

Agent&  $p$    & PO  &  $p_{c}$ &   $p'$   & EP\\
\hline \hline

1    &[0.4 0.6]& 1st& [0.5 0.5]& [0.4 0.6]&   $0.0201$\\

\hline

2    &[0.7 0.3]& 2nd & [0.4 0.6]& [0.7 0.3]&   $0.1838$\\

\hline

\end{tabular}
\end{center}

\end{table}

%\vspace{-6mm}

%Now, assume that outcome A was realized. The market maker will pay
%agent 1 $[ln(0.4) - ln(0.5)]$, and will pay agent 2 $[ln(0.7) -
%ln(0.4)]$, with a net payment of $[ln(0.7) - ln(0.5)]$.

\section{SRMs Truthfulness}

Unfortunately, existing prediction markets don't even achieve
truthfulness in ex-post incentive compatibility (Definition 3), and
they operate under restrictive and unrealistic conditions to achieve
truthfulness. In SRMs literature and prediction markets in general,
it is stated that an agent will be truthful (Definition 1) under
myopic participation, i.e., ``It is optimal for traders to report
their true beliefs provided that they ignore the impact of their
reports on the profit they might garner from the future trades ...
~\cite{Chen2009}". Other studies (e.g.,~\cite{Bluffing}) discuss
truthfulness conditions more explicitly. Given our belief model, we
will require three separate conditions in order to analyze and
develop strategy-proof prediction markets.

{\bf Definition 4.} {\it The SRM truthfulness conditions are:

\begin{enumerate}

\item {\bf Predefined Participation Order (PPO):} Agents participate in the market in a predefined order.

\item {\bf One Participation and Market Influence (OP-MI):} Each agent
participates only once in the market, and can influence the public
signals of other agents only through its single report (i.e., any
external communication (e.g., discussions) is not allowed).

\item {\bf Non-Negative Influence (NNI):} Considering the public signal $p^{pub}$ doesn't decrease the agent's expected payoff.

\end{enumerate}
}

The first two conditions are easy to understand, but the NNI
condition needs more elaboration. Consider the probability
distribution $p^{nat}$ defined by ``nature" over the set $\Omega$ of
all possible outcomes for the future event, where $p^{nat}$ dictates
the ``real" probability that each outcome $i \in \Omega$ will be
realized. Considering any arbitrary market current estimate $p_{c}$
in any SRM, the absolute (or global) maximum expected payoff any
agent can ever get in principle is attained by changing $p_{c}$ to
$p^{nat}$. No agent knows the estimate $p^{nat}$ for certain, but
each agent thinks that its true belief $p = M(p^{prv}, p^{pub})$ is
equal to $p^{nat}$ or at least hopes that $p$ is very close to
$p^{nat}$, since the closer $p$ is to $p^{nat}$ the higher the
agent's expected payoff will be. Given a proper scoring rule $s$,
this can be expressed by the discrepancy value $D(s, p, p^{nat})$,
where the smaller the discrepancy value the closer is $p$ to
$p^{nat}$. The NNI condition means that each agent believes that
when using its public signal $p^{pub}$ to produce its true belief
$p$, the true belief $p$ is closer to $p^{nat}$ than $p^{prv}$,
i.e., $D(s, p, p^{nat}) \leq D(s, p^{prv}, p^{nat})$. This means
that using the public signal $p^{pub}$ to produce the final belief
$p$ has a non-negative influence on (i.e., doesn't decrease) the
agent's expected payoff, as $p^{pub}$ enhances the agent's private
belief $p^{prv}$ about the future event. We stress that the NNI
condition does not necessarily hold if all agents are truthful
(Definition 1), e.g., an agent may report its true belief, but this
belief will negatively influence other agents. That's why we state
the NNI condition -- whether other agents are truthful or not --
rather than stating an ex-post incentive compatibility condition
(Definition 3).

To show that the conditions in Definition 4 are necessary and
sufficient for agents in SRMs to be truthful (Definition 1), we will
show that by dismissing each condition separately, an agent can
maximize its expected payoff by manipulating the market (Definition
2). Then, we will show that if these three conditions hold
simultaneously, then an agent has no incentive to manipulate the
market.

{\bf Theorem 1.} {\it A SRM without the PPO condition is manipulable
even if the OP-MI and NNI conditions hold.}

{\it Proof.} We prove this theorem using a
counter-example~\footnote{It is sufficient to show that an agent has
incentive to manipulate the market given a particular private type
of another agent, without assuming that the agent knows the private
type of that other agent.} that shows that a SRM is manipulable even
when agents participate only once, report their true beliefs and we
neglect any influences between the agents' beliefs (i.e., OP-MI and
NNI conditions hold). Consider Example 1, and assume that agents
there participated in a predefined order.

%\vspace{-4mm}

\begin{table}[!h]
\begin{center}
\caption{Example 2.} \label{tab2} \footnotesize
\begin{tabular}{|c|c|c|c|c|c|}

\hline

Agent&  $p$    & PO  &  $p_{c}$ &   $p'$   & EP\\
\hline \hline

2    &[0.7 0.3]& 1st& [0.5 0.5]& [0.7 0.3]&   $0.0823$\\

\hline

1    &[0.4 0.6]& 2nd & [0.7 0.3]& [0.4 0.6]&   $0.1920$\\

\hline

\end{tabular}
\end{center}
\end{table}

%\vspace{-5mm}

In Example 2 (Table~\ref{tab2}), we assume that agent $1$ has the
chance to report after agent $2$. First, agent $2$ changes $p_{c} =
p_{0} =$ [0.5 0.5] to $p' =$ [0.7 0.3], then agent $1$ changes
$p_{c} =$ [0.7 0.3] to $p' =$ [0.4 0.6]. Agent 1 expected payoff
will be $0.1920$, which is higher than its expected payoff (i.e.,
$0.0201$ in Table~\ref{tab1}) when it participated before agent 2.
$\square$

%Agent 2 uses the merging function $M'$ of Example 1 where the final
%belief $p$ is $p^{prv}$ after increasing the probability of the
%outcome that had a higher probability in $p^{pub}$ by $0.1$, and
%decreasing the the probability of the other outcome by $0.1$.

{\bf Theorem 2.} {\it A SRM without the OP-MI condition is
manipulable even if the PPO and NNI conditions hold.}

{\it Proof.} We prove this theorem by using a counter-example
(recall Footnote 4) that shows that a SRM is manipulable when agents
participate in a predefined order (i.e., PPO condition holds) and
when agents believe that considering their public signals doesn't
decrease their expected payoff (i.e., NNI condition holds). Using
example 2, agent 1 has -- after considering its private and public
signals -- a true belief of [0.4 0.6] and will report after agent 2
according to a predefined order. Agent 2 has a private signal
$p^{prv} =$ [0.7 0.3], and considers the market current estimate as
its public signal $p^{pub}$. Agent 2 uses the merging function $M'$
of Example 1. As shown in Table~\ref{tab2}, the true belief $p$ of
agent 2 will be its private signal $p^{prv} =$ [0.7 0.3], because
the market current estimate $p_{c} = p_{0} =$ [0.5 0.5] is the
public signal and it doesn't favor any outcome.

%\vspace{-3mm}

\begin{table}[!h]
\begin{center}
\caption{Example 3.} \label{tab3} \footnotesize
\begin{tabular}{|c|c|c|c|c|c|}

\hline

Agent&  $p$    & PO  &  $p_{c}$ &   $p'$   & EP\\
\hline \hline

1    &[0.4 0.6]& 1st& [0.5 0.5]& [0.51 0.49]&   $-0.0042$\\

\hline

2    &[0.8 0.2]& 2nd & [0.51 0.49]& [0.8 0.2]&   $0.1809$\\

\hline

1    &[0.4 0.6]& 3rd & [0.8 0.2]& [0.4 0.6]&   $0.3819$\\

\hline

\end{tabular}
\end{center}
\end{table}

%\vspace{-2mm}

In Example 3 (Table~\ref{tab3}), agent 1 participates twice and
makes the first and third reports according to a predefined order.
Because agent 1 believes that its belief $p =$ [0.4 0.6] is true,
then it will try -- from its point of view -- to maximize its
expected payoff by manipulating the public signal of agent 2 as
follows. Agent $1$ will first change $p_{c} = p_{0} =$ [0.5 0.5] to
[0.51 0.49]. Now the market current estimate is [0.51 0.49], which
is the public signal of agent $2$ and it favors outcome $A$ with
probability $0.51$. According to the merging function $M'$ of agent
$2$, its final true belief will be its private signal $p^{prv} =$
[0.7 0.3] after adding $0.1$ to the probability of outcome $A$ and
subtracting $0.1$ from the probability of outcome $B$, i.e., [(0.7 +
0.1) (0.3 - 0.1)] = [0.8 0.2]. After agent 2 reports its belief [0.8
0.2], agent $1$ will change the market current estimate $p_{c} =$
[0.8 0.2] to its true belief [0.4 0.6]. The net expected payoff of
agent 1 from its first and second reports is $-0.0042 + 0.3819 =
0.3777$, which is higher than its expected
payoff~\footnote{Misleading other agents is not always beneficial,
and an agent must balance its loss when bluffing (e.g., agent 1
negative expected payoff from its first false report) with its
future gains.} $0.1920$ when reporting only once after agent $2$ as
in Example 2 (Table~\ref{tab2}). Similar to Example 3, an agent can
gain from affecting the public signals of other agents through
communications outside the market. $\square$

%Agent $1$ can first change $p_{c} = p_{0} =$ [0.5 0.5] to [0.51
%0.49], and its expected payoff from that report will be $0.4
%[ln(0.51) - ln(0.5)] + 0.6 [ln(0.49) - ln(0.5)] = -0.0042$.

%However, agent $1$ can first change $p_{c} =
%p_{0} = [0.5 0.5]$ to $[0.3 0.7]$, and its expected
%payoff from that report will be $0.4 [ln(0.3) - ln(0.5)] + 0.6
%[ln(0.7) - ln(0.5)] = -0.0024$. Then, agent $1$ can change
%$p_{c} = [0.3 0.7]$ to its true belief
%$p^{1} = [0.4 0.6]$, and its expected payoff from
%that report will be $0.4 [ln(0.4) - ln(0.3)] + 0.6 [ln(0.6) -
%ln(0.7)] = 0.0226$.

% Idea, rearrange the agents to achieve a minimum lose ...

{\bf Theorem 3.} {\it A SRM without the NNI condition is manipulable
even if the PPO and OP-MI conditions hold.}

{\it Proof.} Relaxing the NNI condition implies that an agent will
believe that its public signal $p^{pub}$ doesn't enhance its private
belief $p^{prv}$, which means that considering $p^{pub}$ while
formulating its true belief $p = M(p^{prv}, p^{pub})$ may decrease
its expected payoff, i.e., $D(s, p^{prv}, p^{nat}) < D(s, p,
p^{nat})$ may hold. In this case, the agent is better off by
neglecting its public signal $p^{pub}$, and will report its private
belief $p^{prv}$. Reporting $p = p^{prv}$ and not $p = M(p^{prv},
p^{pub})$ violates the agent's truthfulness (Definition 1), and is
considered a strategic misreporting. This can happen under the PPO
and the OP-MI conditions. $\square$

%When agents start neglecting their public signals, they will not
%influence each other's beliefs and this destroys the influence
%factor which is crucial for any prediction market to converge.

%For instance, in the bluffing example we provided in the proof of
%Theorem 2, assume that $w =$ [0.7 0.3]. If agent $2$ reported its
%private signal [0.7 0.3] as its true belief, then this report has a
%discrepancy value of $0$ when compared to $w$. However, if agent $2$
%get affected by the public signal (after agent $1$ false report) and
%reported its final belief [0.8 0.2], then this report has a
%discrepancy value greater than $0$ when compared to $w$.

{\bf Theorem 4.} {\it In a SRM, the PPO, OP-MI and NNI conditions
are necessary and sufficient for agents to be truthful.}

{\it Proof.} In theorems 1, 2 and 3, we showed that relaxing each of
the PPO, OP-MI and NNI conditions separately makes a SRM
manipulable, and thus, they are necessary conditions for achieving
truthfulness. We will now show that they are collectively sufficient
conditions by showing the effect of each condition. {\it NNI
Condition Effect:} an agent will assume that its public signal
$p^{pub}$ doesn't decrease its expected payoff (i.e., $D(s, p,
p^{nat}) \leq D(s, p^{prv}, p^{nat})$) and will use it to produce
its final true belief $p = M(p^{prv}, p^{pub})$. This will hold for
any arbitrary merging function $M$, since the NNI condition
encapsulated the effect of $p^{pub}$ on an agent's expected payoff
regardless of $M$. {\it PPO Condition Effect:} an agent is
participating in a predefined order and will be paid based on
$p_{c}$ that will be fixed according to this predefined order. {\it
OP-MI Condition Effect:} Given any estimate $p_{c}$, the agent can't
influence $p_{c}$ because the agent couldn't have reported it, and
the agent couldn't have affected the public signal of the agent who
reported it. Given the effects of the PPO and OP-MI conditions, the
agent has no control over the estimate $p_{c}$ based on which it
will be paid and $s_{i}(p_{c}), \forall i \in \Omega$ in Eq.4 are
considered constants from the agent's perspective. Given the effect
of the NNI condition and that a SRM uses a proper scoring rule,
reporting $p' = p$ maximizes the agent's expected payoff by
maximizing $\sum^{N}_{i=1} p_{i} s_{i}(p')$. $\square$

%\vspace{-2mm}

\section{Strategy-Proof SRMs}

%\vspace{-3mm}

We will present strategy-proof SRMs that achieve truthfulness in
dominant strategy while dismissing all the previously mentioned
conditions. We will start dismissing one condition at a time to
illustrate the condition's effect on the market maker's WCL.

{\bf Relax PPO.} Consider a SRM with any arbitrary proper scoring
rule $s$, it is easy to see in Eq.~\ref{eq6} that the agent's
expected payoff $\textsl{E}(s, p, p, p_{c})$ (i.e., Eq.~\ref{eq4})
when changing the market current estimate $p_{c}$ to its true belief
(i.e., report $p' = p$) is simply the discrepancy function $D(s, p,
p_{c})$ of the proper scoring rule $s$ (i.e., Eq.~\ref{eq3}).

%\vspace{-6mm}

\begin{eqnarray} \label{eq6}
&\textsl{E}(s, p, p, p_{c}) = \sum^{N}_{i=1} p_{i}
(s_{i}(p) - s_{i}(p_{c})) \\
\nonumber & = \sum^{N}_{i = 1} p_{i} s_{i}(p) - \sum^{N}_{i = 1}
p_{i} s_{i}(p_{c}) = D(s, p, p_{c})
\end{eqnarray}

%\vspace{-1mm}

This implies that in a SRM, an agent's expected payoff is the
discrepancy distance between its report and the previous report made
to the market (i.e., the market's current estimate $p_{c}$). Thus,
an agent can maximize its expected payoff by choosing to go after
the estimate $p_{c}$ that has the greatest discrepancy value
compared to its own report. We present an arbitrary participation
(AP) SRM (Procedure~\ref{Proc2}) that works without the PPO
condition, while an agent will not benefit from delaying its
report~\footnote{We stress that an agent's public signal is
predefined and the agent can't affect the signal's content by
altering its participation time.}.

%\vspace{-2mm}

\begin{algorithm}
\caption{AP SRM Protocol.} \label{Proc2} \algsetup{indent=2em}
\begin{algorithmic}[1]

\STATE $m \leftarrow 1$, $t \leftarrow 0$, $p_{c} \leftarrow p_{0}$
and $\theta(t) \leftarrow p_{c}$.

\REPEAT

\STATE An agent $j$ reports an estimate $p'$ and $p_{c} \leftarrow
p'$.

\STATE $t \leftarrow t + 1$ and $\theta(t) \leftarrow p'$.

\UNTIL{Market closes and outcome $i$ is realized.}

\WHILE{$m \leq t$}

\STATE $\displaystyle p' \leftarrow \theta(m)$, compute $p'_{c} =
\underset{p_{c} \in \theta}{\operatorname{argmax}}[\sum^{N}_{i = 1}
p'_{i} $ $ s_{i}(p') - \sum^{N}_{i = 1} p'_{i} s_{i}(p_{c})]$, and
$m \leftarrow m + 1$.

\STATE The payment of agent $j$ who reported $p'$: $s_{i}(p') -
s_{i}(p'_{c})$.

\ENDWHILE

\end{algorithmic}
\end{algorithm}

%\vspace{-2mm}

{\bf Theorem 5.} {\it Given an AP SRM, an agent will be truthful
under the OP-MI and NNI conditions.}

{\it Proof.} The AP SRM (Procedure~\ref{Proc2}) is similar to a SRM
(Procedure~\ref{Proc1}) except for Step 7. The NNI condition has the
same effect as in theorem 4. The OP-MI condition has the same effect
as in theorem 4, and thus, the agent can't affect the vector
$\theta$ that holds all the reported estimates to the market from
its beginning till its end. In an AP SRM, an agent is paid based on
the estimate $p'_{c}$ that maximizes its expected payoff given its
reported estimate $p'$, where $p'_{c}$ is chosen from vector
$\theta$ and has the greatest discrepancy value compared to $p'$. In
other words, the AP SRM market will pay each agent assuming that it
reported directly after the $p_{c}$ that corresponds to the greatest
discrepancy value compared to the agent's report $p'$. Thus, an
agent has no incentive to delay its participation time. By
substituting $p'_{c}$ in Eq.4, an agent's expected payoff is
$\textsl{E}(s, p, p', p'_{c}) = \sum^{N}_{i=1} p_{i} (s_{i}(p') -
s_{i}(p'_{c}))$, we can consider $s_{i}(p'_{c}), \forall i \in
\Omega$ as the constants that maximize $\textsl{E}(s, p', p',
p'_{c})$. Given that an AP SRM is using a proper scoring rule,
reporting $p' = p$ maximizes the agent's expected payoff by
maximizing $\sum^{N}_{i=1} p_{i} s_{i}(p')$. $\square$

In principle, a SRM (Procedure~\ref{Proc1}) pays and receives
payments from every agent, but because the agents' payments offset
each other (i.e., each agent pays what the previous agent receives),
the SRM pays only the last participating agent. This is not the case
anymore in an AP SRM, and what an agent pays is not what the
previous agent receives because $p'_{c}$ is not necessarily the
previous report made to the market. This implies that we must
consider a payment for each agent, and this increases the market's
WCL by the factor of the number $n$ of participating agents as
follows.

%\vspace{-3mm}

\begin{equation}
n \times \max_{i \in \Omega} \sup_{p, p_{c} \in \Delta_{N}}
(s_{i}(p) - s_{i}(p_{c})).
\end{equation}

%\vspace{-3mm}

However, this is the minimum WCL required to dismiss the PPO
condition as we now show.

{\bf Lemma 6.} {\it The AP SRM is truthful without the PPO condition
and has the minimum WCL.}

{\it Proof.} In a SRM (Procedure~\ref{Proc1}) without the PPO
condition, an agent is free to choose its participation time. The
maximum expected payoff an agent can get by altering its
participation time is by reporting after $p'_{c}$ that has the
greatest discrepancy value compared to its belief $p$. And thus, the
minimum expected payoff that prevents the agent from altering its
participation time is to pay the agent based on $p'_{c}$, which is
the AP SRM payment. $\square$

{\bf Relax OP-MI.} We present a non-myopic (NM) SRM
(Procedure~\ref{Proc3}) which works without the OP-MI condition,
while an agent will not benefit from participating more than once
and/or misleading other agents. The NM SRM is similar to a SRM
(Procedure~\ref{Proc1}) except for Steps 6-9 concerning how agents
are paid, and for Step 3 where agents report their private beliefs
$p^{prv}$ along with their final beliefs $p = M(p^{prv}, p^{pub})$.
However, the market current estimate $p_{c}$ changes according to
the final beliefs. Let $\theta^{prv}$ denote the vector that holds
all reported private estimates.

%\vspace{-2mm}

\begin{algorithm}
\caption{NM SRM Protocol.} \label{Proc3} \algsetup{indent=2em}
\begin{algorithmic}[1]

\STATE $p_{c} \leftarrow p_{0}$.

\REPEAT

\STATE An agent $j$ reports an estimate $p'$, a private estimate
$p^{prv}$ and $p_{c} \leftarrow p'$.

\STATE Add $p^{prv}$ to $\theta^{prv}$.

\UNTIL{Market closes and outcome $i$ is realized.}

\FORALL{Participating Agents}

\STATE Let $p'$ be the last report by agent $j$ made to the market,
and $p^{prv}_{c}$ is the immediate previous private belief from
$\theta^{prv}$ reported to the market before $p'$ by any agent $k
\neq j$ .

\STATE The payment of agent $j$: $s_{i}(p') - s_{i}(p^{prv}_{c})$.

\ENDFOR

\end{algorithmic}
\end{algorithm}

%\vspace{-2mm}

{\bf Theorem 7.} {\it Given a NM SRM, an agent will be truthful
under the PPO and NNI conditions.}

{\it Proof.} The NNI condition has the same effect as in theorem 4.
Without the OP-MI condition, an agent can get the maximum expected
payoff by: {\it 1.} report to the market again and again every time
it realizes that reporting yields a positive expected payoff; and
{\it 2.} by influencing the public signals of other agents for
future gains (e.g., example 3 in theorem 2), either inside the
market by making misleading reports or outside the market. For the
first point, the NM SRM eliminates the incentive to do this by
paying an agent based on its last report to the market (Step 7), and
will neglect all the agent's previous reports. For the second point,
the NM SRM eliminates the incentive to do that by paying the agent
(Steps 7-8) based on the private beliefs of other agent which is not
affected by their public signals. Given the PPO condition and that
agent $j$ is paid based on $p^{prv}_{c}$ that was reported by
another agent $k \neq j$ immediately before the last report made by
agent $j$, an agent can't choose the $p^{prv}_{c}$ which will be
used in its payment. By substituting $p^{prv}_{c}$ in Eq.4, an
agent's expected payoff is $\textsl{E}(s, p, p', p^{prv}_{c}) =
\sum^{N}_{i=1} p_{i} (s_{i}(p') - s_{i}(p^{prv}_{c}))$, and given
the previous, $s_{i}(p^{prv}_{c}), \forall i \in \Omega$ are
considered constants from the agent's perspective. Given that a NM
SRM is using a proper scoring rule, reporting $p' = p$ maximizes the
agent's expected payoff by maximizing $\sum^{N}_{i=1} p_{i}
s_{i}(p')$. $\square$

In NM SRM, the agents' payments don't offset each other, because
each agent is paid by the market according to its report $p'$ (i.e.,
$s_{i}(p')$) and pays the market according to $p^{prv}_{c}$ (i.e.,
$s_{i}(p^{prv}_{c})$), and $p^{prv}_{c}$ is not the $p'$ based on
which the previous agent is paid. This implies that we must consider
a payment for each agent, and this increases the market's WCL by a
factor of the number $n$ of agents (i.e., the same loss as in Eq.6).

{\bf Lemma 8.} {\it The NM SRM is truthful without the OP-MI
condition and has the minimum WCL.}

{\it Proof.} Recalling the two points in the proof of theorem 7
about how an agent can increase its expected payoff, the NM SRM
avoids the first point without an extra loss by paying an agent only
for its last report. For the second point, the NM SRM eliminated any
incentives for an agent from misleading other agents by paying it
according to the private beliefs of other agents. In SRM
(Procedure~\ref{Proc1}) without OP-MI, the maximum gain an agent can
get is by misleading the previous agent to report a $p_{c}$ that has
the greatest discrepancy value for the agent's report $p'$. The
minimum amount to eliminate such an incentive for an agent to
mislead another agent is the WCL of the NM SRM. $\square$

{\bf Relax PPO and OP-MI.} We present an arbitrary participation
non-myopic (AP NM) SRM (Procedure~\ref{Proc4}) which works without
the PPO and the OP-MI conditions by combining the ideas behind the
AP SRM and NM SRM.

{\bf Theorem 9.} {\it Given an AP NM SRM, an agent will be truthful
under the NNI condition.}

{\it Proof.} The proof here is similar to the proof of theorem 7.
The NNI condition has the same effect as in theorem 4, and paying an
agent according to the private beliefs of other agents has the same
effect as in theorem 7. However when we dismiss the PPO condition,
the agent can maximize its expected payoff by altering its
participation time to report $p'$ after the private signal
$p^{prv}_{c}$ that has the greatest discrepancy value from $p'$. But
similar to the AP SRM, the AP NM SRM pays an agent based on
$p''_{c}$ that is chosen from all the reported private estimates
$\theta^{prv}_{\setminus j}$ of other agents and maximizes the
agent's expected value given its report $p'$. Here, the agent has no
incentive to alter its participation time. Given that agent $j$
can't influence the vector $\theta^{prv}_{\setminus j}$, and by
substituting $p''_{c}$ in Eq.4, an agent's expected payoff is
$\textsl{E}(s, p, p', p''_{c}) = \sum^{N}_{i=1} p_{i} (s_{i}(p') -
s_{i}(p''_{c}))$, and we can consider $s_{i}(p''_{c}), \forall i \in
\Omega$ as the constants that maximize $\textsl{E}(s, p, p',
p''_{c})$. Given that an AP NM SRM is using a proper scoring rule,
reporting $p' = p$ maximizes the agent's expected payoff by
maximizing $\sum^{N}_{i=1} p_{i} s_{i}(p')$. $\square$

The WCL in an AP NM SRM is the same as the loss in an AP SRM, but it
pays an agent based on the estimate that maximizes the agent's
expected payoff that is chosen from $\theta^{prv}_{\setminus j}$
rather than $\theta_{j}$ in the AP SRM.

%\vspace{-3mm}

\begin{algorithm}
\caption{AP NM SRM Protocol.} \label{Proc4} \algsetup{indent=2em}
\begin{algorithmic}[1]

\STATE $p_{c} \leftarrow p_{0}$.

\REPEAT

\STATE An agent $j$ reports an estimate $p'$, a private estimate
$p^{prv}$ and $p_{c} \leftarrow p'$.

\STATE Add $p^{prv}$ to $\theta^{prv}$.

\UNTIL{Market closes and outcome $i$ is realized.}

\FORALL{Participating Agents}

\STATE Let $p'$ be the last report of agent $j$ to the market,
compute the vector $\theta^{prv}_{\setminus j}$ from $\theta^{prv}$
to hold the reported private beliefs of all agents except agent $j$,
and compute $p''_{c} = \underset{p^{prv}_{c} \in
\theta^{prv}_{\setminus j}}{\operatorname{argmax}} [\sum^{N}_{i = 1}
p'_{i} s_{i}(p') - \sum^{N}_{i = 1} p'_{i} s_{i}(p^{prv}_{c})] $.

\STATE The payment of agent $j$: $s_{i}(p') - s_{i}(p''_{c})$.

\ENDFOR

\end{algorithmic}
\end{algorithm}

%\vspace{-2mm}

{\bf Lemma 10.} {\it The AP NM SRM is truthful without the PPO and
OP-MI conditions and has the minimum WCL.}

{\it Proof.} Directly follows from Lemmas 6 and 8. $\square$

{\bf Relax PPO, OP-MI and NNI.} Finally, we present a strategy-proof
SRM (Procedure~\ref{Proc5}) that works without any conditions, and
agents will behave truthfully.

{\bf Theorem 11.} {\it Given a strategy-proof SRM, an agent will be
truthful.}

{\it Proof.} Similar to the proof of theorem 7, any agent $j$ is
paid only for its last report, and based on the private signals
reported by other agents, and thus, it has no incentive to
participate more than once and/or manipulate other agents' public
signals either inside or outside the market. When we dismiss the NNI
condition, an agent will not assume that using its public signal
enhances its private signal and may decide to neglect its public
signal. To avoid this, the strategy-proof SRM (Steps 7-8) pays agent
$j$ the maximum amount when outcome $i$ is realized according to
either its reported final belief $p'$ or its reported private signal
$p^{jprv'}$. Thus, the agent doesn't care about the implications of
using its public signal. Agent $j$ is paid based on the private
signals $p^{1}_{c}$ and $p^{2}_{c}$ in $\theta^{prv}_{\setminus j}$
that correspond to the maximum discrepancy value compared to the
agent's reported beliefs $p'$ and $p^{jprv'}$, and thus, the agent
has no incentive to alter its participation time. Given that agent
$j$ can't influence the vector $\theta^{prv}_{\setminus j}$, and
substituting $p^{1}_{c}$ and $p^{2}_{c}$ in Eq.4 results in two
expected payoff equations $\textsl{E}(s, p, p', p^{1}_{c})$ and
$\textsl{E}(s, p^{jprv}, p^{jprv'}, p^{2}_{c})$, agent $j$ considers
$s_{i}(p^{1}_{c})$ and $s_{i}(p^{2}_{c}), \forall i \in \Omega$ in
both equations as the constants that maximize these equations. Given
that the market is using a proper scoring rule, reporting $p' = p$
and $p^{jprv'} = p^{jprv}$ (where $p^{jprv}$ is the true private
signal of agent $j$) maximizes the agent's expected payoff in the
two equations by maximizing $\sum^{N}_{i=1} p_{i} s_{i}(p')$ and by
maximizing $\sum^{N}_{i=1} p^{jprv}_{i} s_{i}(p^{jprv'})$. $\square$

%\vspace{-2mm}

\begin{algorithm}
\caption{Strategy-proof SRM Protocol.} \label{Proc5}
\algsetup{indent=2em}
\begin{algorithmic}[1]

\STATE $p_{c} \leftarrow p_{0}$.

\REPEAT

\STATE An agent $j$ reports an estimate $p'$, a private estimate
$p^{prv}$ and $p_{c} \leftarrow p'$.

\STATE Add $p^{prv}$ to $\theta^{prv}$.

\UNTIL{Market closes and outcome $i$ is realized.}

\FORALL{Participating Agents}

\STATE Let $p'$ be the last report of agent $j$ in the market,
$p^{jprv'}$ be the private belief of agent $j$ reported with $p'$,
compute the vector $\theta^{prv}_{\setminus j}$ from $\theta^{prv}$
to hold the reported private beliefs of all agents except agent $j$,
compute $p^{1}_{c} = \underset{p^{prv}_{c} \in
\theta^{prv}_{\setminus j}}{\operatorname{argmax}} [\sum^{N}_{i = 1}
p'_{i} s_{i}(p') - \sum^{N}_{i = 1} p'_{i} s_{i}(p^{prv}_{c})] $,
and compute $p^{2}_{c} = \underset{p^{prv}_{c} \in
\theta^{prv}_{\setminus j}}{\operatorname{argmax}} [\sum^{N}_{i = 1}
p^{jprv'}_{i} s_{i}(p^{jprv'}) - \sum^{N}_{i = 1} p^{jprv'}_{i}
s_{i}(p^{prv}_{c})]$.

\STATE The payment of agent $j$: $\max [(s_{i}(p') -
s_{i}(p^{1}_{c})) , (s_{i}(p^{jprv'}) - s_{i}(p^{2}_{c}))]$.

\ENDFOR

\end{algorithmic}
\end{algorithm}

%\vspace{-2mm}

Again, the WCL in the strategy-proof SRM is same as that of the AP
SRM and the AP NM SRM.

{\bf Remark 1.} We previously assumed that each agent receives only
one private signal and one public signal. Our work extends to
scenarios where an agent receives its private and public signals and
reports to the market, and then receives new private and public
signals and reports again to the market, as long as there is no
strategic interaction between the two reports (e.g., the agent is
not waiting to receive the new signals).

{\bf Remark 2.} Our work extends to design strategy-proof convex
CFMs. In CFMs~\cite{Chen2007}, the market maker trades a security
for each outcome $i \in \Omega$ that pays \$1 if and only if outcome
$i$ was realized, and \$0 otherwise. Let $q_{i}$ denote the number
of shares of security $i$ held currently by the agents, and $q =
[q_{1}, \ldots, q_{N}]$ denote the vector of shares of all
securities currently held by the agents where $q \in
\mathbb{R}^{N}$. The securities are priced based on a cost function
$C(q): \mathbb{R}^{N} \rightarrow \mathbb{R}$, which describes the
amount of money currently wagered in the market as a function of the
quantity of shares $q$ held by agents. The instantaneous price of
buying an infinitesimal amount of security $i$ is given by $p_{i}(q)
= \partial C(q)/\partial q_{i}$. Let $p(q) = [p_{1}(q), \ldots,
p_{N}(q)]$ denote the vector of prices for all the securities. An
agent trades a bundle  $r = [r_{1}, \ldots, r_{N}] \in
\mathbb{R}^{N}$, where $r_{i}$ is the amount of shares purchased (or
sold if negative) from security $i$. When the agent purchases $r$,
the agent pays the market $C(q + r) - C(q)$. Given an agent's true
belief $p$, the agent's expected payoff from purchasing $r$ is
$\textsl{E}(C, p, q, r) = \sum^{N}_{i=1} p_{i} r_{i} - (C(q + r) -
C(q))$. An agent maximizes its expected payoff by buying a bundle
$r$ such that the prices in the market after this purchase is
equivalent to the agent's true belief, i.e., $p(q+r) = p$.

%Given a cost function $C$, the worst case loss is simply the
%difference between the maximum amount that the market maker might
%have to pay the winners and the amount of money collected by the
%market maker, defined as follows.
%
%\begin{equation}\label{eq13}
%\sup_{q \in \mathbb{R}^{N}} ( sup_{i \in \Omega} q_{i} - (C(q) -
%C(\overrightarrow{0})) )
%\end{equation}

In~\cite[Theorem 3]{Chen2010}, a one-to-one mapping was shown
between a set of convex CFMs and a class of strictly proper SRMs.
This mapping guarantees that an agent who changes the market current
estimate from $p_{c}$ to $p$ in a SRM has exactly the same expected
payoff for every outcome $i \in \Omega$ as the agent who changes the
quantity vector $q$ to $q+r$ such that $p(q) = p_{c}$ and $p(q+r) =
p$ in a convex CFM. This equivalence guarantees that all the
strategic behaviors we indicated for SRMs and the measures taken to
prevent them still hold for convex CFMs. We just need to illustrate
how the extra payments in the strategy-proof SRMs are made in convex
CFMs. In strategy-proof convex CMFs, agents need to report their
private signals along with their purchases. Once an agent purchases
a bundle, the market maker can realize the agent's true belief by
inspecting the market's prices after the purchase. After the market
closes, the market maker can determine at which point of time agent
$j$ would have preferred to make its purchase in order to maximize
its expected payoff based on the private beliefs reported by other
agents. Then, the market maker can issue (i.e., give for free) more
securities to agent $j$ in order to equate its expected payoff from
its purchase with the expected payoff as expressed in any of the
previous procedures. In practice, this will increase the market
maker's loss. The WCL of strategy-proof convex CFMs needs further
investigation since the WCL in traditional SRMs is bounded
irrespective of the number of participating agents and the
equivalence in~\cite{Chen2010} doesn't state that this bound holds
for any convex CFMs. The WCL is bounded for a convex CFM if the
conjugate of its $C(q)$ is bounded over the convex hull of the
probability simplex~\cite[Theorem 3]{Abernethy2011}.

%\subsection{Side Notes}
%
%Only the MSR and DPM are known to guarantee the market's loss is
%bounded.
%
%Idea: If accepting trades that sell if the price is higher and buy
%if the prices are lower, can we bound the lose in cost-based
%function ?
%
%\subsection{Terminology}
%
%Truthful betting: is the strategy of immediately changing the market
%probabilities to the player's probabilities. In other words,

%\section{Discussion and Related Work}

% Strategic behavior in prediction market is well recognized and studied before
% But, up to our knowledge ... such studies aimed for investigating such prediction markets
% from a game-theoretic point of view, i.e., the equilibrium of the game. Not designing new
% prediction markets that have stronger truthfulness properties.

% The increase in the prediction market payments to maintain stronger truthfulness is natural
% and obvious ... this trade off relation between truthfulness and payments is well known in MD

% And more complicated prediction market models shows an o(n) increase in payments as well ...

% Our work can be extended more complicated settings .... the undesirable action model

%\vspace{-3mm}

\section{Conclusions and Future Work}

%\vspace{-2mm}

We showed that designing prediction markets that have the strongest
form of truthfulness (i.e., strategy-proofness) is possible, but
comes with an unbounded WCL for SRMs. However, some commonly used
markets (e.g., CFMs) may have unbounded loss. Moreover, assuming
that markets will operate under very restrictive -- and almost
impossible to hold -- conditions such as PPO, OP-MI or NNI is
unrealistic, e.g., how a market maker will guarantee the order of
the reports made to the market. As well, we have shown that this is
the minimum possible WCL required to dismiss these conditions. This
trade off between the money loss and achieving a strong form of
truthfulness is very common in mechanism design and in prediction
markets with more complicated settings (e.g.,~\cite{Shi2009}).
Extending current ideas to other types of prediction markets appears
fruitful avenue of pursuit.

\section{Acknowledgment}

NICTA is funded by the Australian Government as represented by the Department of Broadband,
Communications and the Digital Economy and the Australian Research
Council through the ICT Centre of Excellence program.

%\vspace{-1.5mm}
%\small
\bibliographystyle{abbrv}
\bibliography{Strategy-Proof_Prediction_Markets}

\end{document}